\documentclass[twocolumn,showpacs,preprintnumbers,amsmath,amssymb]{revtex4}
\usepackage{graphicx}
\usepackage{dcolumn}
\usepackage{bm}

\begin{document}

\title{A theoretical and numerical approach to ``magic angle" of stone skipping }
\author{Shin-ichiro Nagahiro}
 \email{nagahiro@cmpt.phys.tohoku.ac.jp}
\author{Yoshinori Hayakawa}
\affiliation{%
Department of Physics, Tohoku University, Aoba-ku, Sendai, 980-8578, Japan
}%
\date{\today}
\begin{abstract} 
We investigate the condition for the bounce of circular disks which 
obliquely impacts on fluid surface.
An experiment [ Clanet, C., Hersen, F. and Bocquet, 
L., {\it Nature} {\bf 427}, 29 (2004) ] revealed that there exists a 
``magic angle" of $20^\circ$ between a disk's face and water surface
 in which condition the required speed for bounce is minimized.
We perform three-dimensional simulation of the disk-water impact by means of  
the Smoothed Particle Hydrodynamics (SPH). Futhermore, we analyze 
the impact with a model of ordinal differential equation (ODE).
Our simulation is in good agreement with the experiment.
The analysis with the ODE model gives us a theoretical insight
for the ``magic angle" of stone skipping.
\end{abstract}


\maketitle

Problem of impacts and ricochets of solid bodies against water surface 
have been received a considerable amount of attention 
\cite{karman, richard, may, moghisi, mcma2, gaudet}.
In the early stage, the problem was of importance in naval 
engineering concerning the impacts of canon balls on sea-surface \cite{douglas}. 
Investigations then revealed that there exists a maximum angle of incidence $\theta_{\rm max}$ 
for impacts of spheres, above which the rebound does not occur \cite{johnson1}.
Besides, it was empirically found that the $\theta_{\rm max}$ relates to specific
gravity of sphere $\sigma$ as $\theta_{\rm max} = 18/\sqrt{\sigma}$.
This relation was theoretically explained using a simple 
model of an ordinal differential equation (ODE) \cite{johnson1, hutch}. 
In military engineering today, the problem of water impacts may be 
not as important as that of a century ago, however,  recently it 
attracts renewed interest under the studies of locomotion of 
basilisk lizards \cite{mcma1} and stone-skip \cite{bocquet1}.
\begin{figure}[t]
\begin{center}
\includegraphics[width=6.5cm]{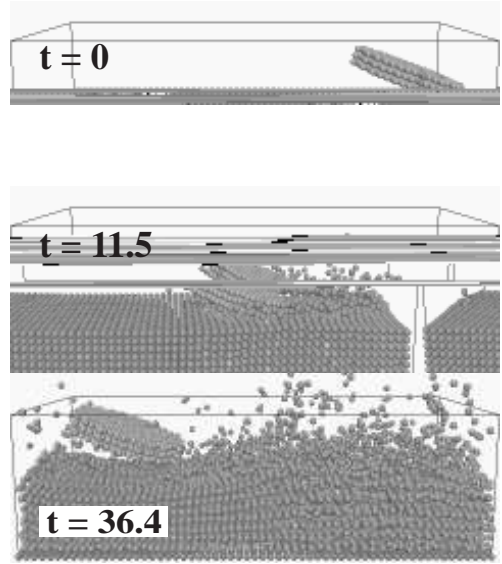}
\caption{Snapshots of the SPH simulation of a disk-water impact with $\phi=20^\circ$ 
and incident angle $\theta=15$. Specific gravity of the disk $\sigma=1.5$ and number of 
fluid particles $N=12600$. }
\label{sph}
\end{center}
\end{figure}

This study is motivated by experimental study
of stone-skip -- bounce of a stone against water surface -- by C. Clanet 
{\it et. al.} \cite{bocquet2}. They investigated impacts of a circular disk 
(stone) on water surface and found that an angle about $\phi = 20^\circ$ 
between the disk's face and water surface would be the 
``magic angle" which minimizes required velocity for bounce. 
In this paper, we study theoretically and numerically 
the oblique impact of disks and water surface. 
Our simulation successfully agrees with the experiment. 
Moreover, we apply an ODE model \cite{birkoff} to the disk-water impact 
and obtain an analytical form of the required velocity $v_{\rm min}$ 
and maximum angle $\theta_{\rm max}$ as a function of initial disk conditions.

To perform a numerical simulation of the disk-water impact, 
we solve the Navier-Stokes equation using the technique of 
Smoothed Particle Hydrodynamics (SPH) \cite{monag, takeda}.
Fig. \ref{sph} is the snapshots of our simulation.
The SPH method is based on Lagrangian description of fluid 
and has an advantage to treat free surface motion.
Several representation of the viscous term have 
been proposed for this method. In this work, we adopt an artificial viscous 
term \cite{cleary1} which is simple for computation and 
sufficiently examined with Couette flow \cite{cleary2}.   
In our simulation, we neglect surface tension and put the 
velocity of sound of the fluid, at least, $25$ times larger than the incident 
velocity of the disk.
 
In the following discussion, we analyze the ODE model 
which was originally introduced by Birkoff et al \cite{birkoff}. 
The model is based on the following assumptions; 
(i) Hydrodynamic pressure $p$ acting from water
is proportional to $({\bf v\cdot n})^2$, where $\bf v$ is the speed of the body and $\bf n$ is the unit vector to the surface of the disk.
(ii) For the part of the surface facing air, there is no hydrodynamic force.
(iii) During the whole process, deformation of water surface is 
negligible, and the boundary between immersed and non-immersed area is 
simply given as the cross section to a horizontal plane at water level.
We notice that the first assumption is reasonable because Reynolds number would 
be of order $10^5$ for typical cases of stone-skip \cite{landau}.

\begin{figure}[tb]
\begin{center}
\includegraphics[height=3.5cm]{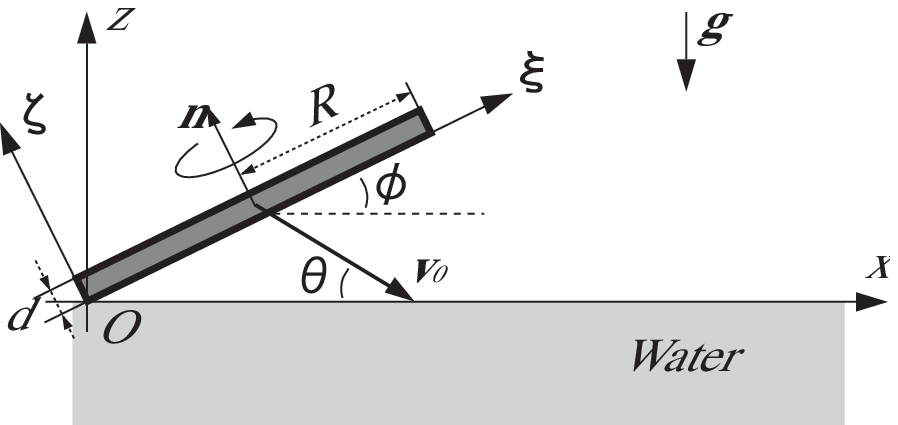}
\caption{An oblique water entry of a circular disk with incident angle $\theta$ and initial velocity ${\bf v}_0$ immediately before the contact (t=0). The edge of the disk is taken at the origin of the coordinates at $t=0$. 
The radius and thickness of the disk is $R$ and $d$. The angle between the disk and water plane is 
$\phi$. }
\label{diskCollView}
\end{center}
\end{figure}

Let us apply the ODE model to the water entry problem with 
circular disk as ``stone" (Fig. \ref{diskCollView}). 
Since the net force ${\bf f}$ to the stone from water is proportional to 
the area $S$ of water-stone interface, we have
\begin{equation}
 {\bf f} = -\frac{1}{2}C_{\rm D}S\rho ({\bf v\cdot n})^2{\bf n},
\end{equation}
where $\rho$ is the density of water, $\bf n$ normal vector.
We need to mention that the drag coefficient $C_{\rm D}$ is 
not necessarily constant during the impact. For example, it 
varies according to the depth of immersion for vertical entries 
of spheres or disks \cite{moghisi,mcma2}. 
Hence in the present case, $C_{\rm D}$ would also depend on conditions such as 
initial angles $\theta$, $\phi$ and time. However, unfortunately, there is no 
experimental data available to determine $C_{\rm D}$ to our purpose. In this 
study, we use $C_{\rm D}\sim 1.4$, which is obtained from our SPH 
simulations of the typical initial angles ($\theta=20^{\circ}, 
\phi=20^{\circ}$) in the experiment (\cite{bocquet2}) and assume $C_{\rm D}$ is constant throughout 
the impact process. 

For simplicity, we limit ourselves to the case that angular velocity
of the disk along the axis $\bf{n}$ is large enough, so that the angle $\phi$ remains constant during the process owing to a gyroscopic effect \cite{bocquet1}. Both the experiments and our 
SPH simulations support the validity of this simplification. 

Taking a frame of reference $O$-$\xi\zeta$ as shown in Fig. 
\ref{diskCollView}, we write the equation of motion as
\begin{eqnarray}
\ddot{\xi} &=& -\frac{1}{F}\sin\phi, \label{eqOfMoX}\\
\ddot{\zeta} &=&\frac{C_D\lambda}{2\pi\sigma}
\dot{\zeta}^2S(z)-\frac{1}{F}\cos\phi.\label{eqOfMoZ}
\end{eqnarray}
Here $\xi$, $\zeta$ and $z$ are the position of the lower 
edge of the disk in each coordinates, $\sigma$ is the specific 
gravity, $\lambda=R/d$, and $F=v_0^2/gR$ is Froude number.

These equations give us a straight forward insight for
a necessary condition for stone-skip. Because the acceleration along
with $\xi$ direction is constant all the time, starting from the initial conditions with
\begin{equation}
\theta + \phi \geq \pi/2,
\end{equation}
one can find the stone always depart downward from the water surface,
therefore, skip of stone would not take place.  
However, this argument does not account the seal of air cavity behind 
the disk. Our SPH simulation gives $\theta+\phi\lesssim0.87\pi/2$ 
for stone-skip domain in the condition of no gravitation
 ($F\to\infty$).

Let us consider the lowest velocity for a bounce $v_{\rm min}$ as a function of the angles $\theta$ and $\phi$. A straight forward criterion
whether a stone skips or not would be the vertical position after
the entry into water surface; if the stone could recover the position higher than the water level, one could say it skips ( criterion A ).
However, to make the analysis simple, we adopt an alternative criterion ( criterion B ); 
if the velocity $\dot{z}$ of the disk changes its sign
to positive we regard ricochet takes place. Under this definition, the 
entry velocity such that the trajectory of the disk have an inflection
point on its horizontal line would give the minimum velocity 
$v_{\rm min}$ for ricochet. 

We can derive an equation which describes trajectories of disk
motion. Eq. (\ref{eqOfMoX}) could be easily integrated with initial conditions $\xi(0)=0$ and $\dot{\xi}(0) = \cos(\theta+\phi)$.
Using the expression of $\xi(t)$, one could replace the time 
derivative in Eq. (\ref{eqOfMoZ}) with that of $\xi$ and obtain
\begin{eqnarray}
&&\left\{ \cos^2(\theta+\phi)-\frac{2\sin\phi}{F}\xi\right\}\zeta''
-\frac{\sin\phi}{F}\zeta' \nonumber\\
&&= \frac{C_D\lambda}{2\pi\sigma}\left\{ \cos^2(\theta+\phi)-\frac{2\sin\phi}{F}\xi\right\}{\zeta'}^2S(z)\nonumber\\
&&~~-\frac{1}{F}\cos\phi, \label{eqorbit}
\end{eqnarray}
where the prime indicates derivative with $\xi$. 

Assume now that the disk entered into water at the minimum 
velocity $v_{\rm min}$. When the disk
reaches to the inflection point $(\xi^*, \zeta^*)$,
$\zeta'=-\tan\phi$ and $\zeta''=0$, we have  
\begin{eqnarray}
\frac{C_D\lambda}{2\pi\sigma} \left\{ \cos^2(\theta+\phi)-\frac{2\sin\phi}{F_{\rm min}}\xi^* \right\}S(z^*)\tan^2\phi\nonumber\\
-\frac{1}{F_{\rm min}\cos\phi}= 0. \label{flexpoint}
\end{eqnarray}
where $F_{\rm min}={v_{\rm min}}^2/gR$. In order for the criterion B to be satisfied, it is necessary that the inflection point exists in the domain of $\xi^*>0$ 
and $z^*<0$. It turns out that in Eq. (\ref{flexpoint}) $\xi^*$
has the maximum value $\hat{\xi}^*$ when the disk is fully immersed, 
{\it i.e.,} $S(z^*)=\pi$. Solving Eq. (\ref{flexpoint}) for
$F_{\rm min}(v_{\rm min})$, we finally obtain an expression for 
$v_{\rm min}$ as
\begin{eqnarray}
v_{\rm min}=\frac{\sqrt{2gR}}{\cos(\theta+\phi)}
	\left\{\hat{\xi}^*\sin\phi + \frac{\sigma\cos\phi}{C_{\rm D}
	\lambda\sin^2\phi}
	\right\}^{\frac{1}{2}},\label{vmin}
\end{eqnarray}
We could derive the critical incident 
angle $\theta_{\rm cr}$ in the same way. Solving Eq. (\ref{flexpoint}) for $\theta$,
\begin{eqnarray}
\theta_{\rm cr} = \arccos{\sqrt{\frac{2}{F}\left(\hat{\xi}^*\sin\phi+\frac{\sigma\cos\phi}{C_{\rm D}\lambda \sin^2\phi}\right) }}-\phi.\label{thetamax}
\end{eqnarray}
Note that, in the limit $F\to\infty$, this equation again gives 
$\theta_{\rm cr}+\phi = \pi/2$.

A position of the inflection point $\hat{\xi}^*$ still remains unknown. 
We treat $\hat{\xi}^*$ as a fitting parameter, which should be determined 
so as to agree with experiments.
However we cannot make a direct comparison between the Eq. (\ref{vmin})
and the experimental data (\cite{bocquet2}) because $v_{\rm min}$ and 
$\theta_{\rm max}$ are acquired under the criterion A in the experiment. 
We thus fit the Eq. (\ref{vmin}) with the result of 
our SPH simulations performed under criterion B and 
evaluate $\hat{\xi}^*=2.6$. 
Due to the nature of the criterion B, these analytical expressions 
for $v_{\rm min}$ and $\theta_{\rm max}$ should give a lower and upper limit of the stone-skip 
domain respectively.
\begin{figure}[t]
\begin{center}
\includegraphics[height=12cm]{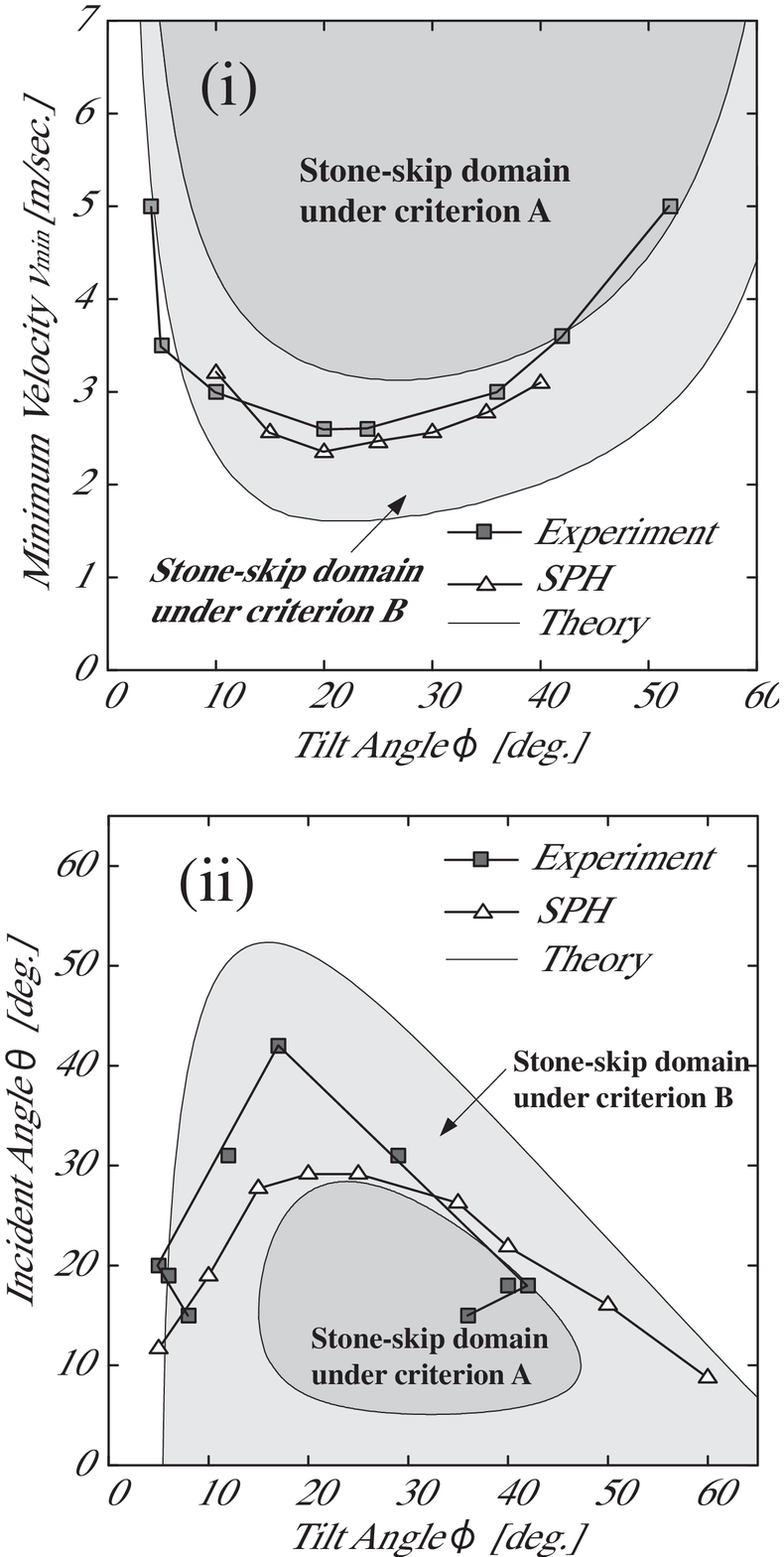}
\caption{ A comparison of the stone-skip domains obtained from 
the experiment (\cite{bocquet2}), SPH simulations and our theory.  
(i), The stone-skip domain in the ($v, \phi$) plane for a fixed $\theta=20^\circ$. 
(ii), The stone-skip domain in the ($\theta, \phi$) plane for a fixed $v=3.5$[m/s]. 
The boundary of the stone-skip domain under the criterion A in each graph are 
numerically drawn and those of B  are the plot of Eqs.(\ref{vmin}) and (\ref{thetamax}) 
respectively. }
\label{ode}
\end{center}
\end{figure}

Then let us discuss our results. We chose the same 
parameters as that of the experiment: $\lambda=9.1$ and $\sigma = 2.7$ unless particularly mentioned. 
Froude number $F$ typically ranges from $4.0$ to $200$. 
For the SPH simulation, $\lambda = 2.5$ and angular velocity 
of the disk $\omega=65[{\rm rounds/s}]$.
Figure \ref{ode} shows the domains of stone-skip in ($\phi, v$) and ($\theta, \phi$) 
planes. For the minimum velocity $v_{\rm min}$, the SPH simulation
successfully agrees with the experiment, and the theoretical results under criterion 
A and B also show the qualitative agreement.
\begin{figure}[t]
\begin{center}
\includegraphics[width=5.5cm]{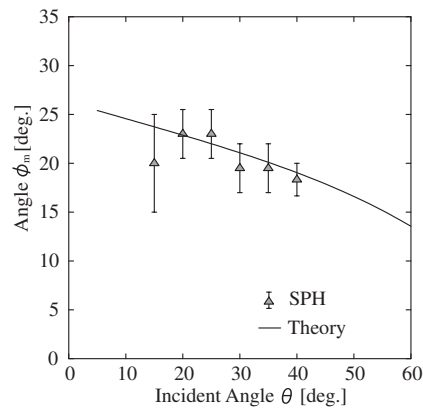}
\caption{Relation of incident angle $\theta$ and the angle $\phi^*$. 
The SPH simulation is performed with $\sigma=2.0$. 
The solid line is obtained numerically 
seeking the minimum of Eq. (\ref{vmin}).}
\label{criticalangle}
\end{center}
\end{figure}

The experiment indicates that the stone-skip domain 
shrinks at $\theta<20^\circ$ in $(\theta, \phi)$ plane. 
The theoretical curve under criterion B does not reproduce 
this tendency while that of criterion A shows the similar 
behavior.  This inconsistency is due to the assumption that the disk is fully immersed 
in the water when it reaches to the inflection point. However, in the case that 
the $\theta$ is much smaller relative to the tilt angle $\phi$ this 
is totally incorrect: only small part of the disk is immersed during the impact 
process. The SPH simulation also shows the different behavior with the 
experiment under $\theta<20^\circ$. We cannot present a clear explanation
for this discrepancy. As for SPH simulations, we mention that, 
the depth of immersion of the disk would be of the order of the fluid particle size of SPH 
at very small incident angle. The numerical error
hence becomes larger for small $\theta$ and for the domain $\theta<10^\circ$
simulation is not attainable.  

The angle $\phi\simeq20^\circ$ is a characteristic 
for both ($\phi, v$) and ($\theta, \phi$) planes in the experiment. 
C. Clanet {\it et. al.} hence suggested that the angle $\phi=20^\circ$ 
is the "magic angle" for stone-skip.
However Eq. (\ref{vmin}) implies that $\phi$ depends on $\theta$. 
In Fig. \ref{criticalangle}, we show how the ``magic angle" $\phi_{\rm m}$
is affected by the incident angle $\theta$.
Our theory suggests $\phi_{\rm m}$ decreases as incident angle increases  
and SPH simulation also shows a decreasing tendency.
However, the change in $\phi_{\rm m}$ is sufficiently
small: $\phi_{\rm m}$ changes only about $15\%$ relative to the change of incident angle 
under $\theta=40^\circ$. 
We therefore conclude that the "magic angle" still 
remains around $\phi=20^\circ$ for the ordinal incident angle at 
stone skipping.

We thank T. Hondou and H. Kuninaka for their helpful suggestions. 
We also acknowledge G. Sakurai and J. Otsuki for valuable discussions. 
This study is supported by the Grant-in-Aid for Scientific 
Study ( Grant No. 1552081 ) from MEXT, Japan.




\end{document}